\documentclass[10pt,conference]{IEEEtran}
\IEEEoverridecommandlockouts

 \pdfoutput=1 

\usepackage{graphics} 
\usepackage{epsfig} 
\usepackage{amsmath} 
\usepackage{amssymb} 
\usepackage{hyperref}
\usepackage{flushend}
\usepackage{caption}
\allowdisplaybreaks
\usepackage{subcaption}
\usepackage[noadjust]{cite}
    
\allowdisplaybreaks

\newcommand{\figwidth}{0.9\columnwidth}
\newcommand{\subfigwidth}{0.9\columnwidth}

\newcommand{\vect}[1]{\boldsymbol{#1}}
\makeatletter
\newcommand{\biggg}{\bBigg@{3}}
\newcommand{\vast}{\bBigg@{4}}
\newcommand{\Vast}{\bBigg@{5}}
\makeatother

\usepackage{standalone}

\usepackage{setspace}
\usepackage{tikz}
\usetikzlibrary{positioning,calc}
\usetikzlibrary{arrows}
\usetikzlibrary{decorations.markings}

\begin{document}

\title{Density Evolution Analysis of Spatially Coupled LDPC Codes Over BIAWGN Channel}

\author{
\IEEEauthorblockN{Md. Noor-A-Rahim, Gottfried Lechner and Khoa D. Nguyen\\}
\IEEEauthorblockA{Institute for Telecommunications Research\\
University of South Australia\\
Adelaide, Australia\\
noomy004@mymail.unisa.edu.au, \{gottfried.lechner, khoa.nguyen\}@unisa.edu.au}
\thanks{This work was supported by the Australian Research Council Grant DE12010016.}
}

%

\maketitle

\begin{abstract}
In this paper, we study the density evolution analysis of spatially coupled low-density parity-check (SC-LDPC) codes over binary input additive white Gaussian noise (BIAWGN) channels under the belief propagation (BP) decoding algorithm. Using reciprocal channel approximation and Gaussian approximation, we propose averaging techniques for the density evolution of SC-LDPC codes over BIAWGN channels. We show that the proposed techniques can closely predict the decoding threshold while offering reduced complexity compared to the existing multi-edge-type density evolution.
\end{abstract}

\begin{IEEEkeywords}
Spatially coupled low-density parity-check codes, density evolution, decoding threshold.
\end{IEEEkeywords}

\IEEEpeerreviewmaketitle
\section{Introduction}
Convolutional LDPC codes or SC-LDPC codes were first introduced by Felstrom and Zigangirov \cite{LDPCC}. In \cite{scldpc}, SC-LDPC codes were analytically investigated over the binary erasure channel (BEC) and it was shown that they exhibit a BP decoding threshold that approaches the maximum-a-posteriori decoding threshold of the underlying ensemble. Over binary memoryless symmetric channels, this \emph{threshold saturation} phenomenon was numerically shown in \cite{SC_BMS,SC_Uni,SC_Kumar,SC_Yadla}. In general, density evolution is used to determine BP decoding threshold of SC-LDPC codes.

Over the BEC, density evolution analysis of SC-LDPC codes is straightforward, since it tracks the erasure probability, which is a scalar quantity. However, this is not the case for the BIAWGN channel. The tracking parameter for the BIAWGN channel is the probability density function (pdf) of the log likelihood ratio (LLR). Thus, the actual density evolution of SC-LDPC codes over BIAWGN channel requires tracking multiple pdfs of messages travelling over multiple edge-types. This results in substantially high computational requirements. To reduce  complexity, the reciprocal channel approximation (RCA) was used within the multi-edge-type (MET)  framework  in \cite{Divas_RCA}, where the tracking parameter of the density evolution analysis is a scalar quantity. 

Due to the MET framework, the density evolution analysis in \cite{Divas_RCA} tracks the parameter of each edge-type separately. Moreover,   deterministic connection between variable node and check node was assumed in  \cite{Divas_RCA}. On the other hand, for SC-LDPC codes with probabilistic coupling between protographs, applying the MET framework is prohibitively complex. Examples of such cases are anytime spatially coupled codes proposed in \cite{Allerton_paper}.

In this paper, we propose a low complexity density evolution approximation for SC-LDPC codes over BIAWGN channel. Together with the RCA and Gaussian approximation (GA), we introduce averaging techniques  in density evolution. Instead of tracking the parameters of RCA or GA on different edge-types, we track  these parameters for different node-types. Since the number of node-types is generally much smaller than the number of edge-types,  the proposed techniques have significantly lower complexity than density evolution with RCA or GA in the MET framework.  The decoding thresholds obtained from the proposed techniques are compared with the exact decoding thresholds obtained from the actual density evolution over the MET framework.  Although the proposed techniques slightly reduce the accuracy, their computational complexity is close to that of density evolution over the BEC.  Moreover, the proposed density evolution techniques are also applicable for spatially coupled codes with random coupling.

The remainder of this paper is organized as follows. We briefly present the background of SC-LDPC codes and  MET framework in Section~\ref{sec:preli}. We present the MET density evolution with approximations in Section~\ref{sec:MET_App}. In Section~\ref{sec:Avg}, we present the proposed  density evolution analysis and show the numerical results in Section~\ref{sec:results}.

\section{SC-LDPC Codes and MET Framework}\label{sec:preli}
SC-LDPC codes \cite{scldpc} are constructed by coupling a chain of $C_L$ standard $(d_v,d_c)$-regular LDPC codes or protographs, where $d_v$ and $d_c$ are the degrees of variable nodes and check nodes, respectively. We refer to $C_L$ as the chain length and assume that the protographs are at positions $[1,2,...,C_L]$, $C_L \in \mathbb{N}$\footnote{In \cite{scldpc}, the authors assumed that the protographs are at positions $[-C_L,C_L]$, $C_L \in \mathbb{N}$.}. Each of the $d_v$ connections of a variable node at position $i$ is connected to their neighbouring check nodes in the range $i$ to $i+\gamma-1$, where $\gamma$ is the coupling length. A SC-LDPC code can be terminated by adding extra check nodes at positions $C_L+1$ to $C_L+\gamma-1$. We refer to this code as $(d_v,d_c,\gamma, C_L)$-SC-LDPC code with design rate 
\begin{align}
R_d = 1-\frac{\frac{d_{v}}{d_{c}}\left(C_L+\gamma+1-2\sum\limits_{i=0}^{\gamma}\left(\frac{i}{\gamma}\right)^{d_{c}}\right)}{C_L}.\nonumber
\end{align}
For simplicity of notations, in the remainder of this paper, we will describe density evolutions for the SC-LDPC codes described in \cite{scldpc}. The proposed scheme can be extended to more general spatially coupled codes shown in~\cite{DE_Thres}. 

\begin{figure}[htbp]
 \centering
\tikzstyle{cir}=[circle, draw, fill=black!100,  minimum width=.3 cm,  minimum height=.3 cm] 
\tikzstyle{block}=[rectangle, draw, fill=black!100,  minimum width=.6 cm,  minimum height=.3 cm]

\begin{tikzpicture}
\node [cir] (c1) at (0,0) {};
\node [cir, below = 2.2 cm of c1] (c2) {};
\node [cir, right = 1.2 cm of c1] (c3) {};
\node [cir, below = 2.2 cm of c3] (c4) {};

\node [block, above= 0cm of {$(c1)!0.5!(c2)$}] (p1){};
\node [block, above= 0cm of {$(c3)!0.5!(c4)$}] (p2){};
\node [block, right = 1.2 cm of p2] (p3) {};
\node [block, right = 1.2 cm of p3] (p4) {};

\draw [-][thick]  (c1)--(p1.north) node [pos = .5, left] {$1$}; 
\draw [-][thick]  (c1)--(p2.north) node [pos = .4, left] {$2$}; 
\draw [-][thick]  (c1)--(p3.north) node [pos = .2, above] {$3$}; 

\draw [-][thick]  (c2)--(p1.south) node [pos = .5, left] {$1$};
\draw [-][thick]  (c2)--(p2.south) node [pos = .4, left] {$2$};
\draw [-][thick]  (c2)--(p3.south) node [pos = .2, below] {$3$};

\draw [-][thick]  (c3)--(p2.north) node [pos = .75, right] {$4$};
\draw [-][thick]  (c3)--(p3.north) node [pos = .95, above] {$5$};
\draw [-][thick]  (c3)--(p4.north) node [pos = .85, above] {$6$};
 
\draw [-][thick]  (c4)--(p2.south) node [pos = .75, right] {$4$};
\draw [-][thick]  (c4)--(p3.south) node [pos = .95, below] {$5$};
\draw [-][thick]  (c4)--(p4.south) node [pos = .85, below] {$6$};



\end{tikzpicture}
\\ 
 \caption{MET representation of  $(d_v,d_c,\gamma, C_L) = (3,6,3,2)$-SC-LDPC code with $m_e = 6$. The edge-types are denoted along side of each edges. }\label{fig:MET_SCLDPC}
\end{figure}
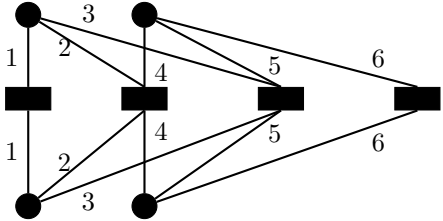

SC-LDPC codes can be included in the MET framework. Using the notation in~\cite[Chapter~5]{mct}, a MET code can be specified by two node perspective multinomials defined as $\vect{L}(\vect{s}) = \sum L_{\vect{q}}\vect{s}^{\vect{q}}$ and $\vect{R}(\vect{s}) = \sum R_{\vect{q}}\vect{s}^{\vect{q}}$, where $\vect{s}$ and $\vect{q}$ are defined by
\begin{itemize}
 \item $\vect{s} = [s_1, ..., s_{m_e}]$ denotes a vector of variables, each corresponds to an edge-type. $m_e$ is the number of edge-types used to represent the graph ensemble;
\item $\vect{q} = [q_1, ..., q_{m_e}]$ denotes degrees of edge-types, where $q_i$ is the number of edges of type $i$ that connect to the same check/variable node.
\end{itemize}
The non-negative coefficient $L_{\vect{q}}$ ($R_{\vect{q}}$) represents the fraction of variable (check) nodes  of type $\vect{q}$. A simple example of a MET SC-LDPC code is shown in Fig.~\ref{fig:MET_SCLDPC}, where we assume deterministic connections between variable and check nodes with $\gamma= d_v$. For this example, the multinomials are given by
\begin{align}
\vect{L}(\vect{s}) &= 0.5s_1s_2s_3+0.5s_4s_5s_6, \nonumber\\
\vect{R}(\vect{s}) &= 0.25s_1^2+0.25s_2^2s_4^2+0.25s_3^2s_5^2 + 0.25s_{6}^2. \nonumber
\end{align}

The MET density evolution is performed by tracking the evolution of pdf of each edge-type message, where the density evolution recursion can be derived from $\vect{L}(\vect{s})$, $\vect{R}(\vect{s})$ and the pdf of the LLR of the received message corresponding to a bit transmitted over the BIAWGN channel  with  noise variance $\sigma_n^2$. For more details about MET density evolution, we refer the readers to \cite[Chapter~5]{mct}. The above density evolution analysis provides an accurate prediction of the decoding threshold of a MET code. However, MET density evolution suffers from high computational complexity, since it tracks the whole distribution of each edge-type message.

\section{MET Density Evolution with Approximation} \label{sec:MET_App}
In this section, we summarise density evolution of SC-LDPC codes using approximations (i.e., RCA and GA) in  the MET framework as depicted in Fig.~\ref{fig:MET_SCLDPC}. These approximations reduce the complexity of the actual MET density evolution analysis by tracking a scalar quantity instead of the whole message distribution along  each edge-type. In \cite{chung2000construction}, it is shown that the message distributions can be well approximated through these scalar quantities and  hence, one can find the threshold without much sacrifice in accuracy. 

Let $\mathcal{N}_v(i)$ be the set of edge-types that share the variable node with an edge of type $i$. Similarly, let $\mathcal{N}_c(i)$ be the set of edge-types that share the check node with an edge of type $i$. Let $x^{(l)}_e(i)$ and $y^{(l)}_e(i)$ $i\in \{1,2,..,m_e\}$ be the scalar quantities, which are deduced from the LLR message distribution through RCA or GA, passed from variable and check nodes, respectively, along the edge-type $i$, at iteration $l$. For RCA, $x^{(l)}_e(i)$ represents the signal-to-noise ratio (SNR) which equals half of the mean of the LLR message, while  $y^{(l)}_e(i)$ represents the sum of the reciprocals of SNRs incoming to the check nodes. For SNR $z$, a reciprocal  $r$ is defined such that $C_f(z) + C_f(r)  = 1$, where $C_f(u)$ denotes the mutual information between the input and the output of a BIAWGN channel with SNR $u$ \cite{Divas_RCA}. For GA,  $x^{(l)}_e(i)$ and $y^{(l)}_e(i)$  represent the mean of the LLR message. The RCA and GA in the MET framework are more rigorously described in the following.


\subsection{The RCA Technique}

\begin{itemize}
\item Initialization: $x^{(0)}_e(i)  = \frac{1}{\sigma_n^2}$.

\item $y_e^{(l)}(i)$ update:
\begin{align}
y^{(l)}_e(i) = &(q_i-1) C_f^{-1}\left(1-C_f \left(x_e^{(l)}(i)\right)\right)+ \nonumber\\
\hspace{1 cm} & \sum\limits_{j\in \mathcal{N}_c(i)} {q_j}   C_f^{-1}\left(1-C_f \left(x_e^{(l)}(j)\right)\right).\nonumber
\end{align}

\item $x^{(l)}_e(i)$ update:
\begin{align}
x^{(l)}_e(i) =& \frac{1}{\sigma_n^2} + (q_i-1) C_f^{-1}\left(1-C_f \left(y_e^{(l-1)}(i)\right)\right) + \nonumber\\
&\quad \quad \sum\limits_{j\in \mathcal{N}_v(i)}  q_jC_f^{-1}\left(1-C_f \left(y_e^{(l-1)}(j)\right)\right).\nonumber
\end{align}
\end{itemize}



\subsection{The GA Technique}

\begin{itemize}
\item Initialization: $x^{(0)}_e(i)  = \frac{2}{\sigma_n^2}$.
\item $y^{(l)}_e(i)$ update:
 \begin{align}
y^{(l)}_e(i) = &\phi^{-1}\Bigg[1-\left(1-\phi \left(x_e^{(l)}(i)\right)\right)^{q_i-1} \nonumber\\
&\hspace{1 cm}.\;\prod\limits_{j\in \mathcal{N}_c(i)} \left(1-\phi \left(x_e^{(l)}(j)\right)\right)^{q_j}\Bigg].\nonumber
\end{align}
\item$x^{(l)}_e(i)$ update:
\begin{align}
x^{(l)}_e(i) = \frac{2}{\sigma_n^2} + (q_i-1) y^{(l-1)}_e(i) +\sum\limits_{j\in \mathcal{N}_v(i)} q_j y^{(l-1)}_e(j),\nonumber
\end{align}
\end{itemize}

where 
\begin{align}
 \phi(u)=
 \begin{cases}
 1 - \frac{1}{\sqrt{4\pi u}}\int_{-\infty}^{\infty} \big(\tanh \frac{f}{2}\big) e^{-\frac{(f-u)^2}{4u}}df, & \text{if} \hspace{0.2 cm} u>0 \\\nonumber
 1, & \text{if} \hspace{0.2 cm} u=0 \nonumber
 \end{cases}
 \end{align} 

\noindent
For both RCA and GA techniques, we find the decoding threshold $\sigma^*_n$ by
\begin{align}
\sigma^*_n= \sup\left\{\sigma_n:\lim_{l\to\infty} x^{(l)}_e(i) = \infty, \quad \forall i\right\}.\nonumber
\end{align}


\section{Proposed Density Evolution Analysis} \label{sec:Avg}
In BEC, an average of the  incoming erasure probabilities  at the variable or check nodes can be used to calculate the outgoing erasure probabilities \cite{scldpc}. This property results in a low complexity density evolution analysis of SC-LDPC codes over the BEC. Motivated by the density evolution over  BEC, we propose averaging techniques in the density evolution analysis over BIAWGN channels, while utilizing RCA (or GA). In our proposed density evolution, we track the  messages from different nodes instead of tracking the  message for different edge-types. We compute the outgoing messages from a node based on an averaging function of the incoming messages.  Using RCA and GA, we propose averaging the mutual information between the incoming message and  the associated code bit.  For each variable or check node, we convert each of the incoming quantity (SNR or mean of LLR message)  to the corresponding mutual information and average these  mutual information. Then, the average mutual information is converted back to the corresponding quantity, which is used to calculate the outgoing quantity.  For the proposed density evolution analysis, we consider $x^{(l)}(i)$ and $y^{(l)}(i)$ to be the scalar quantities\footnote{The nature of $x^{(l)}(i)$ and $y^{(l)}(i)$ are same as  $x^{(l)}_e(i)$ and $y^{(l)}_e(i)$, respectively as mentioned in Section~\ref{sec:MET_App}.}, outgoing from variable and check nodes, respectively. The  indices $i$ and $l$ denote the node position and iteration number, respectively. In the following density evolution, we set $x^{(l)}(i) = \infty$, $\forall l$ when $i\not\in [1,C_L]$.  Using RCA and GA, in the following we present the proposed update steps of $x^{(l)}(i)$ and $y^{(l)}(i)$ for $i\in [1,C_L]$. 


\subsection{The RCA Technique}

\begin{itemize}
\item Initialization: $x^{(0)}(i) = \frac{1}{\sigma_n^2}$.

\item $y^{(l)}(i)$ update:
\begin{align}
y^{(l)}(i) =(d_{c}-1)C_f^{-1}\left(1-\frac{1}{\gamma}\sum\limits_{k=0}^{\gamma-1}  C_f\left(x^{(l)}(i-k)\right) \right). \nonumber
\end{align}


\item $x^{(l)}(i)$ update:
\begin{align}
&x^{(l)}(i) =  \frac{1}{\sigma_n^2} + (d_{v}-1) \nonumber\\
&\quad.\; C_f^{-1}\left(1-\frac{1}{\gamma}\sum\limits_{k=0}^{\gamma-1}  C_f\left(y^{(l-1)}(i+k)\right) \right).  \nonumber
\end{align}


\end{itemize}

\subsection{The GA Technique}

\begin{itemize}

\item Initialization: $x^{(0)}(i) = \frac{2}{\sigma_n^2}$.

\item $y^{(l)}(i)$ update:
\begin{align}
&y^{(l)}(i) = \phi^{-1}\biggg[1-\Bigg\{1- \nonumber\\
&\hspace{.5 cm} \left.\left.\phi\left(J^{-1}\left(\frac{1}{\gamma}\sum\limits_{k=0}^{\gamma-1}J\left(x^{(l)}(i-k)\right)\right)\right)\right\}^{d_c-1}\right].\nonumber
\end{align}


\item $x^{(l)}(i)$ update:
\begin{align}
x^{(l)}(i) = \frac{2}{\sigma_n^2} + (d_v-1)  J^{-1}\left(\frac{1}{\gamma}\sum\limits_{k=0}^{\gamma-1}J\left(y^{(l-1)}(i+k)\right)\right),\nonumber
\end{align}


where 
\begin{align}
 J(u)=
 1 - \frac{1}{\sqrt{4\pi u}}\int_{-\infty}^{\infty} e^{-\frac{(f-u)^2}{4u}}\log_2(1+e^{-f})df. \nonumber
 \end{align}

\end{itemize}

\noindent
For both RCA and GA techniques, the decoding threshold is
\begin{align}
\sigma^*_n= \sup\left\{\sigma_n:\lim_{l\to\infty} x^{(l)}(i) = \infty, \quad \forall i\right\}.\nonumber
\end{align}

\noindent
\textbf{Complexity Comparison:} Our proposed density evolution analysis offers a reduction in complexity compared to the earlier MET techniques by tracking node-type instead of edge-type messages. For SC-LDPC codes, the number of messages to be tracked in our proposed approach is at most $\frac{1}{d_v}$ of that in the MET techniques. For example, with $(d_v,d_c,\gamma=d_v, C_L)$ and $(d_v,d_c,\gamma=2d_v, C_L)$-SC-LDPC codes in the MET framework,  there exist $d_v$ and $2d_v$ types of edges, respectively at each position. Thus, MET techniques track $d_vC_L$ and $2d_vC_L$  messages (variable node to check node), respectively.  On the other hand, for both SC-LDPC codes, there exists  single type of node at each position. Thus, our proposed density evolution tracks only $C_L$ messages.

\begin{figure*}
 \centering
 \begin{subfigure}[htbp] {0.45\textwidth}
\includegraphics[width=\subfigwidth]{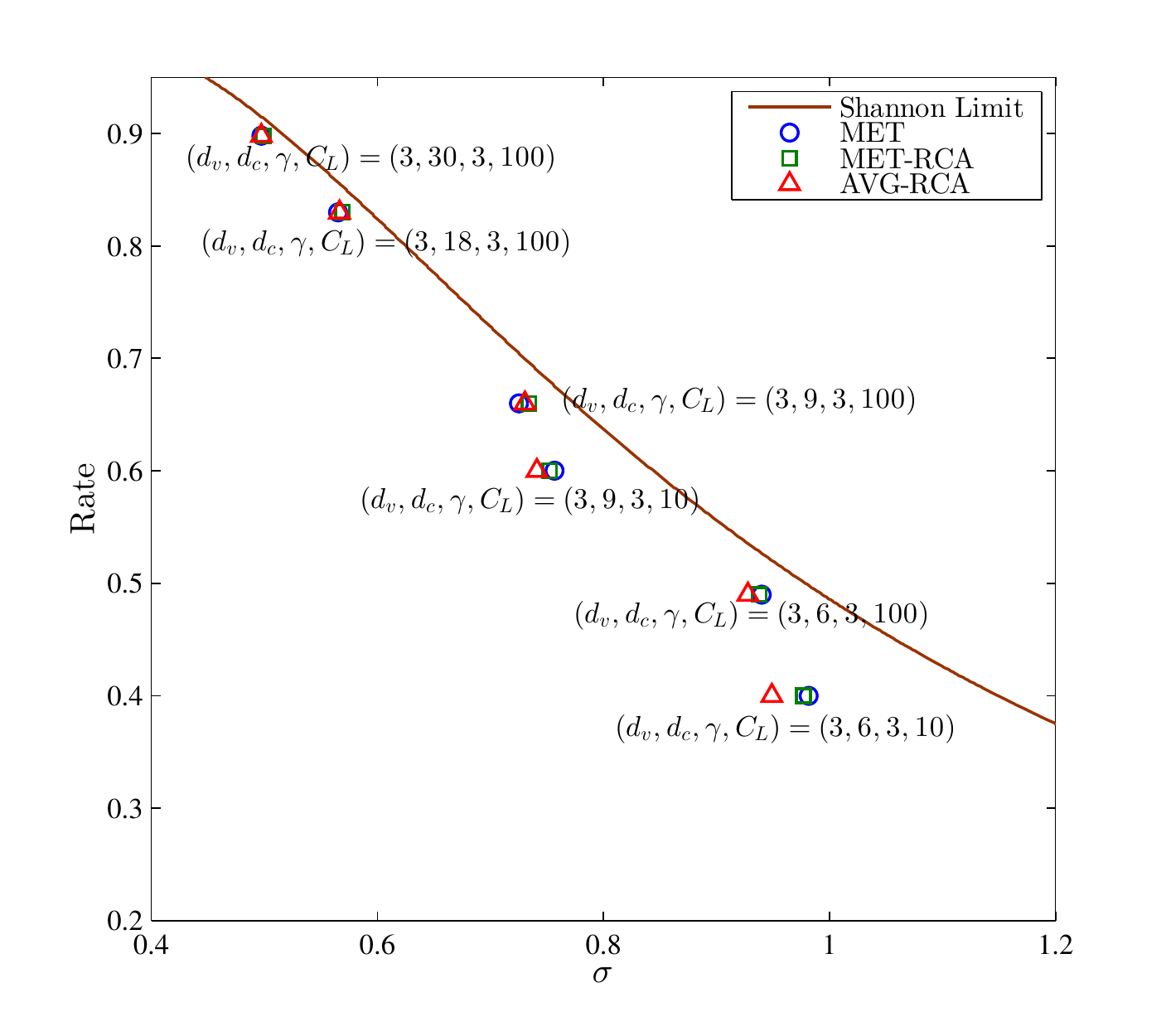}\\
 \caption{With reciprocal channel approximation (RCA) technique}\
 \label{fig:Per_A1_A2}
 \end{subfigure}
 \begin{subfigure}[htbp]{0.45\textwidth}
 \includegraphics[width=\subfigwidth]{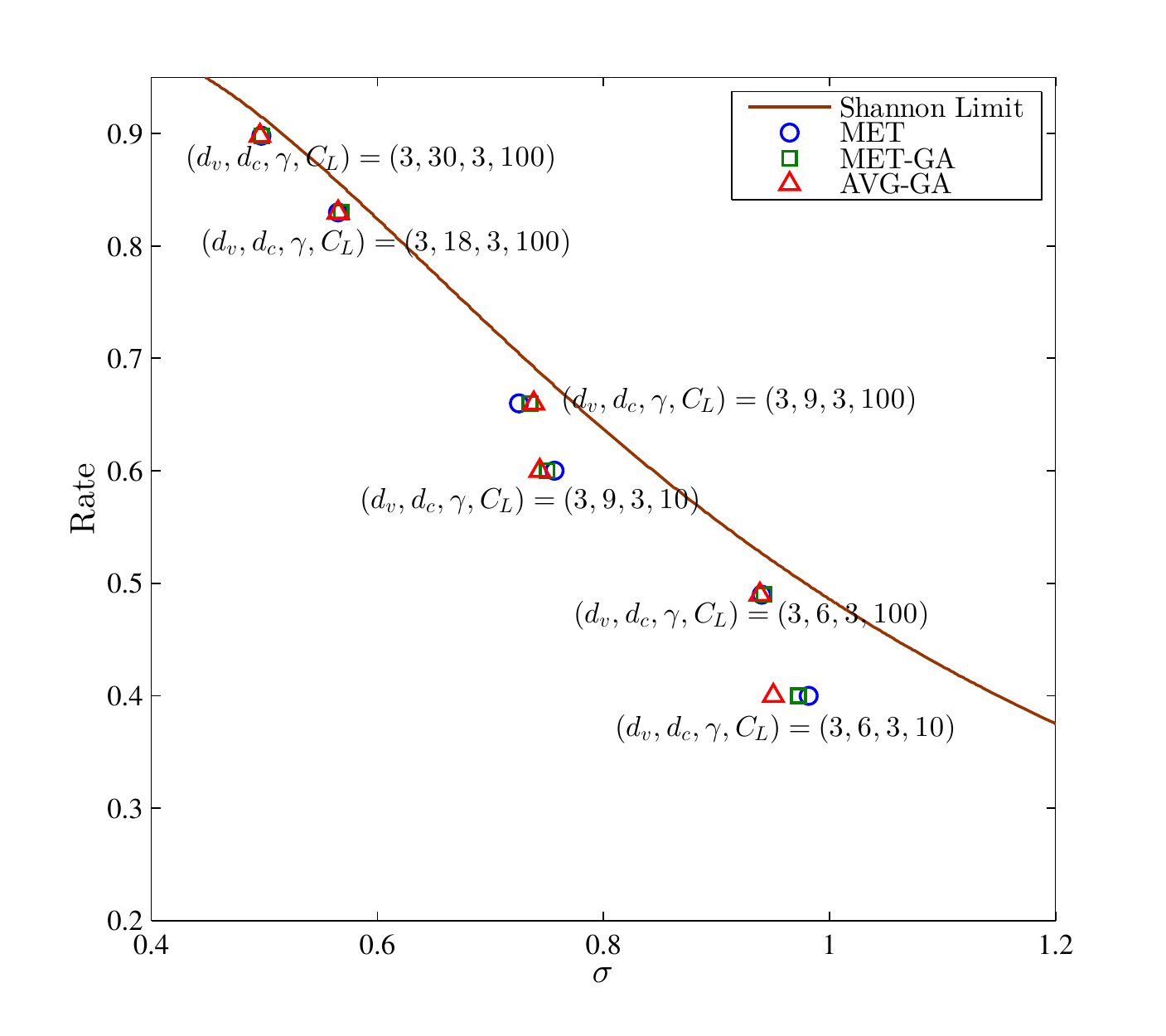}\\
 \caption{With Gaussian approximation (GA) technique}\label{fig:Com_A1_A2}
 \end{subfigure}
 \caption{Decoding thresholds of different SC-LDPC codes over BIAWGN channel.}\label{fig:results}
\end{figure*}

\section{Results and Discussion} \label{sec:results}
In this section, we present numerical results obtained from the aforementioned density evolution techniques.  The thresholds for different SC-LDPC codes  obtained from different density evolution techniques are plotted in Fig.~\ref{fig:results}. We observe that the threshold obtained from RCA-MET and GA-MET density evolutions is a close approximation to the exact threshold obtained from the actual MET density evolution as described in Section~\ref{sec:preli}. Although MET density evolution techniques lead to a good prediction of decoding thresholds, they suffer from high computational complexity and become very complicated when there exist  probabilistic connections between variable and check nodes. 

On the other hand, the decoding thresholds obtained from the proposed RCA-averaging and GA-averaging techniques provide slightly inaccurate approximations, especially at low rates. However, they offer the least computational complexity. Moreover, the proposed techniques provide tractable density evolution analysis for any coupling distribution as well as for probabilistic connections between variable and check nodes. Both RCA and GA averaging techniques compute the decoding threshold with almost same accuracy and complexity. Thus, one can use any of them to measure the decoding threshold of the SC-LDPC codes. It is worth mentioning that the decoding thresholds given by the proposed density evolution are no longer accurate for small chain length $(C_L<5)$ and/or very low rate $(R_d<0.15)$ SC-LDPC codes.

\begin{figure}[htbp]
  \centering
 \includegraphics[width=\figwidth]{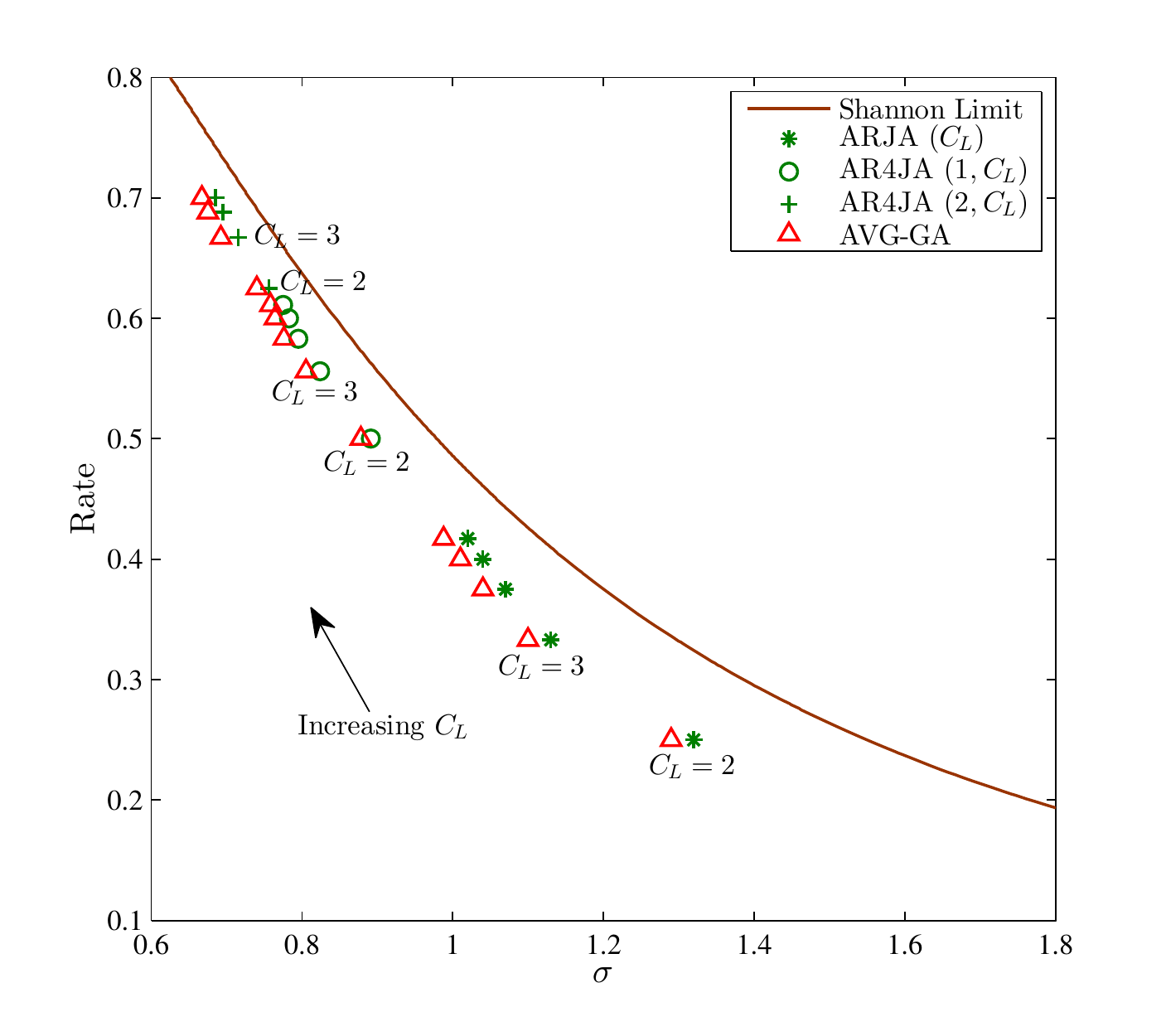}\\
 \caption{Thresholds prediction of spatially coupled ARJA ($C_L$) and spatially coupled AR4JA ($e,C_L$)  codes through proposed averaging technique. In AR4JA ($e,C_L$), $e$ is the extension parameter \cite{DE_Thres}.}\label{fig:ARJA}
\end{figure}

The proposed technique also provides good approximation for the structured protograph-based spatially coupled codes shown in \cite{DE_Thres}.  As an example, we apply the proposed technique for the spatially coupled version of the accumulate-repeat-jagged-accumulate (ARJA) codes and the accumulate-repeat-by-4-jagged-accumulate (AR4JA) codes. The averaging technique for these spatially coupled codes is slightly more complicated than the averaging technique for SC-LDPC codes. Since, there exist multiple types of nodes in an ARJA or AR4JA protographs, one has to track multiple node-type messages from each position. In Fig.~\ref{fig:ARJA}, we show the prediction of actual decoding threshold of these codes using GA-averaging technique.  We  observe that our proposed technique can track the actual thresholds with reasonable accuracy. 


\section{Conclusion}
In this paper, we present density evolution techniques to find the decoding threshold of SC-LDPC codes over BIAWGN channels. We have observed that although MET density evolution results in an accurate analysis, it suffers from high complexity and limited practicality. On the other hand, although our proposed density evolution techniques are slightly less accurate, they are easy to implement and have broader applications.

\bibliography{references}
\bibliographystyle{IEEEtran}
\end{document}